\documentclass[aps,prb,twocolumn,groupedaddress,amsmat,amssymb,amsfonts,superscriptaddress]{revtex4-2}

\usepackage{graphicx}
\usepackage{amsmath,amsfonts,amsthm} 
\usepackage{blkarray}
\usepackage{physics}
\usepackage{mhchem}
\usepackage{caption}
\usepackage{xcolor}
\usepackage{float}
\usepackage{soul}

\newcommand{\nmg}[2]{n_{#1 #2}}
\newcommand{\nmu}[1]{n_{#1 +}}
\newcommand{\nmd}[1]{n_{#1 -}}

\newcommand{\cmg}[2]{c_{#1 #2}^{\dag}}
\newcommand{\cmu}[1]{c_{#1 +}^{\dag}}
\newcommand{\cmd}[1]{c_{#1 -}^{\dag}}

\newcommand{\amg}[2]{c_{#1 #2}}
\newcommand{\amu}[1]{c_{#1 +}}
\newcommand{\amd}[1]{c_{#1 -}}

\begin{document}


\title{Competing multipolar orders in face-centered cubic lattice: application to Osmium double-perovskites}
\author{Derek Churchill}
\affiliation{Department of Physics, University of Toronto, Ontario, Canada M5S 1A7}
\author{Hae-Young Kee}
\email[]{hykee@physics.utoronto.ca}
\affiliation{Department of Physics, University of Toronto, Ontario, Canada M5S 1A7}
\affiliation{Canadian Institute for Advanced Research, CIFAR Program in Quantum Materials, Toronto, Ontario, Canada, M5G 1M1}
\date{\today}

\begin{abstract}
In 5$d^2$ Mott insulators with strong spin-orbit coupling, the lowest pseudospin states form a non-Kramers doublet, which carries quadrupolar and octupolar moments.
A family of double-perovskites where magnetic ions form a face-centered cubic (FCC) lattice, 
was suggested to unveil an octupolar order offering a rare example in d-orbital systems.
The proposed order requires a ferromagnetic (FM) octupolar interaction, since the antiferromagnetic (AFM) Ising model is highly frustrated on the FCC lattice.
A microscopic model was recently derived for various lattices: for an edge sharing octahedra geometry, 
AFM Ising octupolar and bond-dependent quadrupolar interactions were found when only dominant inter- and intra-orbital hopping integrals are taken into account.  
Here we investigate all possible intra- and inter-orbital exchange processes and report that interference of
 two intra-orbital exchanges generates a FM octupolar interaction. 
 Applying the strong-coupling expansion results together with tight binding parameters obtained by density functional theory, 
 we estimate the exchange interactions for the Osmium double-perovskites,  Ba$_2$BOsO$_6$ (B = Mg, Cd, Ca). 
 { Using classical Monte-Carlo simulations, 
 we find that these systems are close to the phase boundary between AFM type-I quadrupole and FM octupole orders. We also find that 
 exchange processes beyond second order perturbation theory including virtual processes via pseudospin-triplet states may stabilize an octupolar order.}
\end{abstract}


\maketitle

\section{Introduction}

In transition-metal Mott insulators, the orbital degeneracy can be lifted by Jahn-Teller effects leading to low energy physics described by a spin-$1/2$ dipole moment.  
However when spin-orbit coupling (SOC) is strong with relatively weak Jahn-Teller coupling, spin and orbital degrees of freedom are entangled and the effective Hamiltonian is described by 
a total angular momentum $J$ often called pseudospin. 
The exchange interactions are determined by the pseudospin wave functions which depend on the number of electrons in d-orbitals.
In general, the entangled-spin-orbit feature is manifested in highly anisotropic exchange interactions leading to rich and novel phenomena in d-orbital Mott insulators.\cite{khaliullin2005orbital,balents2014review,rau2016review,winter2017review,hermanns2018review,Takagi2019NRP,Takayama2021JPSJ}
The most famous example is the Kitaev interaction\cite{kitaev2006anyons} in $d^5$ and $d^7$ with $J_{\rm eff} = 1/2$ wave functions\cite{Jackeli2009PRL}: note
that the wave functions of $d^5$ and $d^7$ are distinct leading to
very different strengths of the bond-dependent $\Gamma$ interaction\cite{Jackeli2009PRL,Rau2014PRL,Liu2018PRB,Liu2020PRL,Sano2018PRB}.
{ Pseudospins can also generate higher-rank multipolar exchange interactions which can compete giving rise to vastly different ground states \cite{Fiore2021PRB,Pouro2019PRB,Pi2014PRB,Pi2014PRL}.
}

{ The $d^2$ pseudospin states form a low energy, non-Kramers $E_g$ doublet and excited $T_{2g}$ triplet (see Fig. 1(a) in \cite{Kha2021PRR}). 
In contrast to the popular $J_{\rm eff}=1/2$ with a dipole moment, the non-Kramers doublet has a vanishing dipole moment similar to $f^2$ ions.
\cite{Fazekas_1999,Chen2010PRB,Chen2011PRB}}
This two-particle spin-orbit-entangled state instead carries quadrupolar and octupole moment, and thus
$d^2$ Mott insulators with strong SOC offer a playground to explore multipolar physics in transition-metal systems.
However, unlike f-electron systems where octupolar orders are extensively investigated\cite{Kubo2006,Santini2009RMP,Kuramoto2009JPSJ,Hotta2012PRI}, 
a long-range octupolar order  in transition metal systems
is highly nontrivial to achieve due to a Jahn-Teller driven orbital order\cite{Kugel1982}.

%
Recently, it was proposed that a family of { insulating,} Osmium (Os) double perovskites exhibits a long-range octupolar order 
offering a first example of octupolar order in d-orbital materials.\cite{Sreekar2020PRB,Maharaj2020PRL,Paramekanti2020PRB}
 The magnetically active Os, hosting a non-Kramers doublet as discussed above, forms a face-centered-cubic (FCC) lattice as shown in Fig. \ref{DPstructure}(a).
Measurements in the heat capacity and magnetic susceptibility for Ba$_2$BOsO$_6$ where B = Mg, Ca show anomalies at approximately $ T^* \sim$ 50 K\cite{Marjerrison2016PRB} suggesting a phase transition. Fitting the suceptibility to the Curie-Weiss law yields a negative Curie-Weiss temperature indicating antiferromagnetic (AFM) interactions\cite{Marjerrison2016PRB}, but neutron diffraction data finds no evidence of any magnetic ordering down to 10 K\cite{Maharaj2020PRL}.
Furthermore, $\mu$SR measurements of the zero field muon decay asymmetry spectra show oscillations implying time-reversal symmetry breaking below approximately 50 K.\cite{Marjerrison2016PRB}
These experimental results suggest they may exhibit octupolar order.\cite{Paramekanti2020PRB, Maharaj2020PRL} 
In particular, octupolar order would provide an explanation for the small field observed using $\mu$SR and absence of detectable dipole order from neutron diffraction experiments\cite{Maharaj2020PRL}.

Motivated by these experimental findings, a microscopic theory was developed for various lattices including double perovskites with different bond geometries.\cite{Kha2021PRR}
It was shown that an intra-orbital exchange process generates bond-dependent quadrupolar interactions, whereas inter-orbital exchanges generate AFM Ising octupolar interactions in addition to ferromagnetic (FM) $xy$-like quadrupolar interaction.\cite{Kha2021PRR} 
Since the AFM Ising interaction is highly frustrated on the FCC lattice, the long-range octupolar order has little room to occur. However, this pioneering work included only the two dominant hopping paths leaving a question on the sign of the octupolar interaction when other hopping paths are included.

In this paper, we investigate all the intra- and inter-orbital exchange processes allowed and show that {\it interference of  two intra-orbital} exchanges 
generates a FM octupolar interaction, which dominates over the AFM contribution from the inter-orbital exchange processes. 
Using these results together with the tight binding parameters obtained by first principle
{\it ab-initio} calculations on Ba$_2$BOsO$_6$ {(B=Mg, Cd, Ca)}, we estimate the multipolar exchange parameters { using a strong coupling perturbation theory.}

{ Within second order perturbation theory, we show that these compounds are close to the boundary between AFM type-I quadrupole and FM octupole.}
{ When virtual processes via the triplet states are included in fourth order perturbation theory, the AFM quadrupolar exchange integral is suppressed. Due to the frusturation of AFM interactions on the 
FCC lattice, FM octupolar ordering is found in one of the systems.}

The paper is organized as follows. In Sec. II, we review the local atomic physics of $5d^2$ and how the non-Kramers doublet {  and excited triplet arise}. 
We then present a nearest-neighbour (n.n.) tight-binding Hamiltonian based on symmetry restrictions. 
Using a strong coupling expansion, we determine the pseudospin Hamiltonian which includes FM Ising octupolar term from
the two paths of  intra-orbital exchange processes. In Section III, we use density functional theory (DFT) to find the tight-binding parameters of {Ba$_2$BOsO$_6$ (B = Mg, Cd, Ca)} and estimate the strengths of the exchange interactions. 
Using classical Monte Carlo simulations, we show { a finite temperature phase diagram for a given set of exchange parameters. We also show a zero temperature phase diagram as a function of quadrupolar and octupolar exchange interactions to show 
how close the systems are to the phase boundary.}
We summarize our results and 
discuss implications of our theory and the possibility of octupolar order in the last section. 

\section{Model Derivation}

In this section, we derive the microscopic spin 
exchange parameters for $d^2$ double-perovskites with strong SOC in an ideal FCC structure.
Before we proceed to the strong coupling expansion to determine the exchange parameters, 
we review the local physics of an isolated  Os atom. Since it is local physics, this can be used as a starting point of
any $d^2$ systems as shown in the earlier work \cite{Kha2021PRR}.

\subsection{Local Physics}

\begin{figure}
    \includegraphics[width=0.42\textwidth]{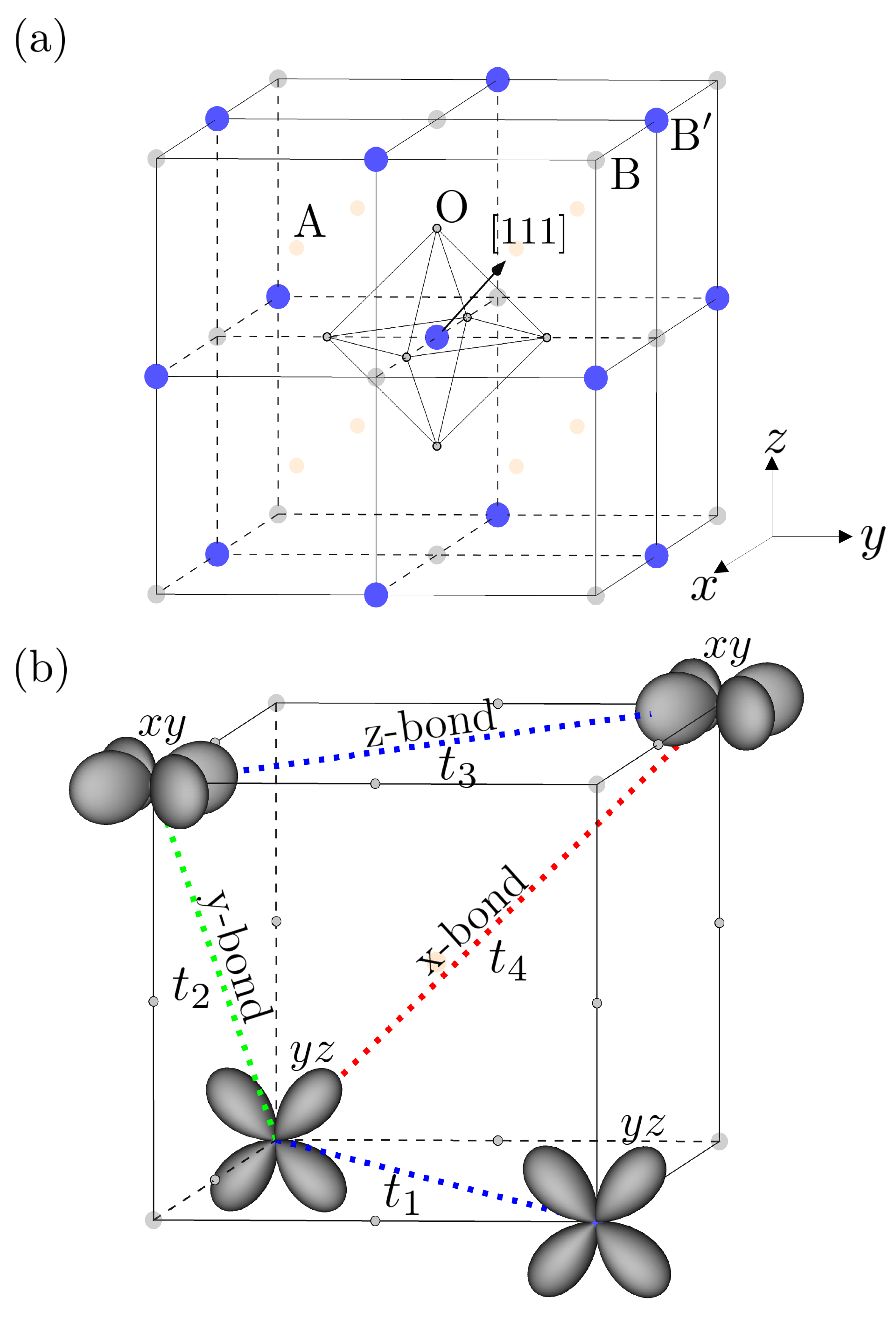}
    \captionof{figure}{(a) Double perovksite (A$_2$BB$'$O$_6$) crystal structure. A atoms are light orange, B atoms are light gray, B$'$ atoms are blue, and oxygen atoms are 
    outlined with black. All oxygen atoms surrounding the B and B$'$ atoms (besides those belonging to the central octahedra) are omitted for clarity. 
    (b) Examples of exchange process among $t_{\rm 2g}$ orbitals. x-, y- and z-bonds are denoted by the red, green and blue 
    dotted lines respectively.
    \label{DPstructure}} 
\end{figure}

First, we discuss the atomic physics for an isolated Os atom. Since each Os atom is surrounded by an oxygen octahedral cage, the $L=2$ irreducible representation
splits into $e_\textrm{g}$ and $t_{\textrm{2g}}$, which are separated by an octahedra crystal field splitting, $\Delta$. 
The local Kanamori-Hubbard Hamiltonian is written  as follows:
\begin{align} \label{kanamoriHam}
    H_\textrm{int} &= U\sum_{m}\nmu{m}\nmd{m} + U'\sum_{m\neq m'}\nmu{m}\nmd{m'} \\
    \notag          &+ \left(U' - J_H\right) \sum_{m < m', \sigma} \nmg{m}{\sigma} \nmg{m'}{\sigma} \\
    \notag & + J_H \sum_{m \neq m'}\cmu{m}\cmd{m'}\amd{m}\amu{m'} \\
    \notag &+ J_H \sum_{m \neq m'}\cmu{m}\cmd{m}\amd{m'} \amu{m'} - \lambda {\bf L} \cdot {\bf S},
\end{align}
where $\cmg{m}{\sigma}$ creates an electron with orbital $m$ and spin S=1/2 denoted by $\sigma=\pm$. 
$U$  and $U' (= U - 2J_H)$ are intra- and inter-orbital Coulomb interactions
respectively, $J_H$ is Hund's coupling, ${\bf L} (= \sum_i {\bf l}_i)$ and ${\bf S} (=\sum_i {\bf s}_i)$ are total orbital angular  and spin momentum respectively, and
SOC $\lambda = \frac{\xi}{2S}$ where $\xi$ is single-particle SOC, i.e, $\sum_i \; \xi \; {\bf l}_i \cdot {\bf s}_i$\cite{Fazekas_1999}. 

The energy hierarchy we will be considering for the above parameters is $\Delta, U > \xi, J_H$ (see Fig. 1 in Ref. \cite{Kha2021PRR}). 
When $H_{\rm int}$ is projected onto the $t_\textrm{2g}$ subspace and restricted to the $n = 2$ sector, we find a $J=2$ ground state for an isolated Os atom. 
As shown in Ref.  \cite{Sreekar2020PRB,Maharaj2020PRL,Paramekanti2020PRB}, taking into account the $e_\textrm{g}$
orbitals, spin-orbit coupling mixes the $t_\textrm{2g}$ and $e_\textrm{g}$ orbitals, which splits the $J=2$ ground state into a non-Kramers $E_{\textrm{g}}$ doublet and an excited triplet.
We use the notation $E_\textrm{g}$ to distinguish  it from $e_\textrm{g}$-orbitals, $d_{x^2-y^2}$ and $d_{3z^2-r^2}$. 
The splitting between the $E_\textrm{g}$ doublet and excited triplet is described by the cubic crystal field Hamiltonian given by
\begin{align} \label{cubicStev}
    H_{\Delta_c} = \Delta_c \left(O^0_4+5O^4_4\right),
\end{align}
where $O^0_4$ and $O^4_4$ are Steven's operators \cite{Maharaj2020PRL,Paramekanti2020PRB}. 
The resulting non-Kramers doublet using $\ket{J_z}$ states is given by 
    \begin{eqnarray}
        \ket{\uparrow} &=& \frac{1}{\sqrt{2}}\left(\ket{-2} + \ket{2}\right) \label{psuedoup},\nonumber\\
        \ket{\downarrow} &= & \ket{0} \label{pseudodown}.
    \end{eqnarray}
$\ket{\uparrow}$ and $\ket{\downarrow}$ are introduced to represent the $E_g$ wavefunctions. 
Since they are either an equal mixture of $|J_z = \pm 2\rangle$ or $|J_z = 0\rangle$, they do not carry a dipole moment, and thus should be differentiated from 
pure spin $\sigma = \pm$ in Eq. 1.

Expressing them in terms of total spin and orbital angular momentum states, $\ket{L_z,S_z}$ is useful, because one can notice
$\ket{\downarrow}$ is elongated in the octahedral z direction whereas $\ket{\uparrow}$ is more flattened in the xy plane (see Fig. 1 in Ref. \cite{Kha2021PRR}); 
these vastly different shapes give rise to interesting features in the effective psuedo-spin model.
    \begin{align}
        \ket{\uparrow} & =  \frac{1}{\sqrt{2}}\left(\ket{1,1} + \ket{-1,-1}\right), \nonumber\\
        \ket{\downarrow} & = \frac{1}{\sqrt{6}}\left(\ket{1,-1} + 2\ket{0,0} + \ket{-1,1}\right).
    \end{align}
Furthermore, we note that the quadrupole ($Q_{x^2-y^2} = J_x^2 - J_y^2$ and $Q_{3z^2} = \left(3J_z^2 - J^2\right) / \sqrt{3}$ \cite{Chen2010PRB}) operators and octupole operator ($T_{xyz} = \frac{\sqrt{15}}{6}\overline{J_x J_y J_z}$ \cite{Chen2010PRB}) form the Pauli matrices
of pseudospin-1/2 operators:
    \begin{gather}
        s_x \equiv \frac{1}{4\sqrt{3}}Q_{x^2-y^2}, \nonumber\\
        s_{y} \equiv \frac{1}{6 \sqrt{5}}T_{xyz}, \nonumber\\ 
        s_z \equiv \frac{1}{4\sqrt{3}}Q_{3z^2},
    \end{gather}
where $s_x$ acting on the pseudo-spin state follows how Pauli matrices typically act on pure spin-$\frac{1}{2}$ states. For example, $s_x \ket{\uparrow} = \frac{1}{2}\ket{\downarrow}$ and $s_x \ket{\downarrow} =\frac{1}{2} \ket{\uparrow}$.
It is important to note that this pseudospin coordinate system is defined in such a way that $s_y$ is along the body-diagonal of the FCC lattice, i.e., [111]-axis shown in Fig. \ref{DPstructure}(a).
Thus the quadrupolar moments lie within the [111]-plane while the octupolar moment is perpendicular to this plane and parallel to [111]-axis.

\subsection{Tight-binding Hamiltonian}
Double-perovskites are a fascinating and rich family of materials exhibiting a variety of magnetic properties \cite{VASALA20151,Chen2010PRB,Marjerrison2016PRB,Hirai2020PRR,Thompson_2014,Nilsen2021PRB,deVries2010PRL,Mustonen2018,Wakabayashi2019,Zhao2021PRB}.
They have the general chemical form A$_2$BB$'$O$_6$ where A belongs to the family of rare-earth elements or alkaline earth metals, B/B$'$ typically belong to the transition metals and O is oxygen. 
The A atoms exist between the B and B$'$ layers and form a cubic lattice, and the oxygens form an octahedral cage around each B and B$'$ atom as shown in  Fig. \ref{DPstructure}(a). \par

In an ideal double perovskites, the B and B$'$ atoms form a pair of interlocking FCC sublattices which can also be viewed as stacked checkerboards of B/B$'$ atoms.
This provides a natural route to geometric frustration and can lead to important consequences on the observed phases.
For Ba$_2$BOsO$_6$ with B = Ca, Mg, Cd, B atoms are non-magnetic leading to a FCC lattice of $d^2$ doublets.

In this subsection, we present the tight-binding Hamiltonian which will be used as a perturbation in the strong coupling expansion later on.
The n.n. tight-binding Hamiltonian between two Os sites on the z-bond is given by
\begin{equation}\label{tbmat_simp}
    t_{ij} = 
    \begin{blockarray}{ccccc} 
     & & \amg{j,xy}{} & \amg{j,xz}{} & \amg{j,yz}{}  \\
    \begin{block}{cc(ccc)}
        \cmg{i,xy}{} & & t_3 & t_4 & t_4 \\
        \cmg{i,xz}{} & & t_4 & t_1 & t_2 \\
        \cmg{i,yz}{} & & t_4 & t_2 & t_1\\
    \end{block}
    \end{blockarray}
    \end{equation}
where $t_i \in \mathbb{R}$. The $C_2$ axis along the [$\overline{1}$10] direction, inversion symmetry about the bond center, and time-reversal symmetry have all been used to restrict the form of this Hamiltonian\cite{Rau2014PRL}. 
This bond will be referred to as a z-bond since $t_3$ is the largest hopping integral and describes the effective overlap of $d_{xy}$ orbitals on n.n. $B'$ sites as displayed in Fig. \ref{DPstructure}(b). 
Under trigonal distortions along the [111] direction (or other distortions where the $C_2$ axis along the bond direction is broken), $t_4$ will be finite. { However, for DPs of interest maintain the $C_2$ axis along the bond direction 
which forces $t_4=0$ due to the symmetry.}
A representative hopping integral of $t_i$ ($i= 1 - 4$) on x, y and z-bonds is shown in Fig. \ref{DPstructure}(b). Note that $t_2$ between $d_{xz}$ and $d_{yz}$ on z-bond
is the hopping between $d_{xy}$ and $d_{yz}$ on the y-bond, indicating the bond-dependence of orbital overlaps
which in turn leads to bond-dependent pseudospin exchange interactions as presented below. 

\subsection{Pseudospin Model}

To derive the effective Hamiltonian for the z-bond, we perform a strong coupling expansion assuming
 that the energy scale of the tight-binding parameters are smaller than the Kanamori interactions. 
The resulting effective Hamiltonian in the ground state formed by the onsite doublets can be found by evaluating
\begin{align}
    \langle \psi_i|&H_{i,j}|\psi_j \rangle =   \sum_{n \notin GS} \frac{\left<\psi_i|{t_{i,j} + t_{i,j}^{\dagger}}|n\right> \left< n|{t_{i,j} + t_{i,j}^{\dagger}}|\psi_j \right>}{E_n - E_0},
\end{align}
where $\ket{\psi_i}$ are the ground states, $E_0$ is the ground state energy, $n$ sums over all excited states, and $t_{ij}$ is the tight-binding Hamiltonian for the z-bond (Eq. \ref{tbmat_simp}) \cite{Mila2010}. 

The resulting effective Hamiltonian for the z-bond is given by
\begin{align}
    H^{z}_{ij} = J_\tau^{(2)} s_{i,z} s_{j,z} + J_q^{(2)} \left(s_{i,x} s_{j,x} + s_{i,z} s_{j,z} \right) + J_o^{(2)} s_{i,y} s_{j,y}, \label{xy_ham}
\end{align}
where 
    \begin{eqnarray}
        J_\tau^{(2)} &=& \frac{4}{9U}\left(t_1 - t_3\right)^2 \label{jt}, \nonumber\\
        J_q^{(2)} &=& \frac{2}{3U} \left[ t_1 \left(t_1 + 2t_3\right) - t_2^2  \right] \label{jq}, \nonumber\\
        J_o^{(2)} &= & \frac{2}{3U} \left[  t_1 \left(t_1 + 2t_3\right) + t_2^2  \right]  \label{jo}.
    \end{eqnarray}
    \label{exchange}
where $J_q^{(2)}$ and $J_o^{(2)}$ contain the product of two intra-orbital hopping integrals, $(t_1  t_3)$ which is negative, as $t_1$ and $t_3$ come in opposite signs,
and dominates over the other terms. 
Here, we have set $J_H =0$ since these are much smaller than U. The effect of finite $J_H$ on the exchange parameters is shown in the next section. 
{ We denote exchange integrals obtained through second order perturbation theory by a superscript (2).}
{ The corrected exchange integrals including virtual triplet processes are shown in Appendix. A, and are denoted by a superscript (4) (see Eq. \ref{exintcor}) and are included in Table 1.}

There are three unique effective Hamiltonians for the 12 n.n bonds which can be obtained by applying $C_3$ rotations about the [111] direction to Eq. \ref{xy_ham}. Under a counter-clockwise $C_3$ rotation, the pseudo-spin operators transform according to
    \begin{gather}
        s_{x} \to \frac{-1}{2}s_{x} - \frac{\sqrt{3}}{2}s_{z} \label{sxtrans}, \nonumber\\
        s_{y} \to s_{y} \label{sytrans}, \nonumber\\
        s_{z} \to \frac{\sqrt{3}}{2}s_{x} - \frac{1}{2}s_{z} \label{sztrans}.
    \end{gather}

Applying these transformations to Eq. \ref{xy_ham} generates terms like $s_{i,x}s_{j,z}$ in the x- and y-bond Hamiltonians. To write the total Hamiltonian compactly, we thus introduce the following operator
\begin{align}
    \tau_i^\gamma = \textrm{cos(}\phi_\gamma\textrm{)}s_{i,z} + \textrm{sin(}\phi_\gamma\textrm{)}s_{i,x}, \label{jtaudef}
\end{align}
where $\gamma \in \{{\rm z,x,y}\}$ referring to three different bonds as shown in the blue, red, green dotted lines in Fig. \ref{DPstructure}(b), 
and their corresponding angle $\phi_{\rm z,x,y} = 0, \frac{2 \pi}{3},\frac{4 \pi}{3}$.

Therefore, the full effective Hamiltonian is given by 
\begin{align}
    H^{\gamma}_{ij} = J_\tau^{(2)} \tau_{i}^\gamma \tau_{j}^\gamma + J_q^{(2)} \left(s_{i,x} s_{j,x} + s_{i,z} s_{j,z} \right) + J_o^{(2)} s_{i,y} s_{j,y}. \label{tot_ham}
\end{align}

Fundamentally, bond dependence of the quadrupolar interactions originates from the vastly different shapes of the doublet wavefunctions; namely, $\ket{\uparrow}$ is flattened in the xy-plane whereas $\ket{\downarrow}$ is stretched in the octahedral z-direction\cite{Kha2021PRR}. 
These differences generate the $J_\tau$ term in Eq. \ref{xy_ham}.
Interestingly, there is also interference between the $t_1$ and $t_3$ hopping processes as evident by the $t_1 t_3$ terms in Eq. \ref{jo} as mentioned above.
The origin of these terms can be traced back to $\ket{\downarrow}$ containing terms like $\cmu{yz}\cmd{xz}$ and $\cmd{yz}\cmu{xz}$ when written in the spin-orbital basis, which are absent in $\ket{\uparrow}$. These terms allow for a combination of $t_1$ and $t_3$ virtual processes to mix $\ket{\uparrow}$ with $\ket{\downarrow}$ and $\ket{\downarrow}$ with $\ket{\downarrow}$, 
generating finite $t_1 t_3$ terms responsible for an FM octupolar interaction. 
Overall intra-orbital hopping $t_1$ and $t_3$ gives rise to bond-dependent quadrupolar and bond-independent octupolar interactions, 
whereas inter-orbital hopping $t_2$ gives rise to only bond-independent interaction.
This is in contrast to $d^5$ systems with $J_{\rm eff} = 1/2$ where intra-orbital hopping $t_2$ is essential for Kitaev interaction\cite{Jackeli2009PRL}, while
the interference of intra- and inter-orbital exchange $t_2 t_3$ leads to $\Gamma$ interaction.\cite{Rau2014PRL} 


It is worthwhile to note that without $t_1$, we have $J_o^{(2)} = -J_q^{(2)}$ and Eq. \ref{tot_ham} reduces to the result shown in Ref. \cite{Kha2021PRR}. 
Moreover, the bond-independent octupolar interactions without $t_1$ are AFM. This results in purely quadrupolar order 
since the AFM octupolar interaction is frusturated on the FCC lattice.  However, the introduction of $t_1$ now causes $J_o^{(2)} \neq J_q^{(2)}$, and also results in FM octupolar interactions (Section IV) which may allow for the possibility of octupolar order. This will be presented after we show the tight binding parameters obtained by DFT.

\section{Density functional Theory}

\begin{figure} 
    \includegraphics[width=0.3924\textwidth]{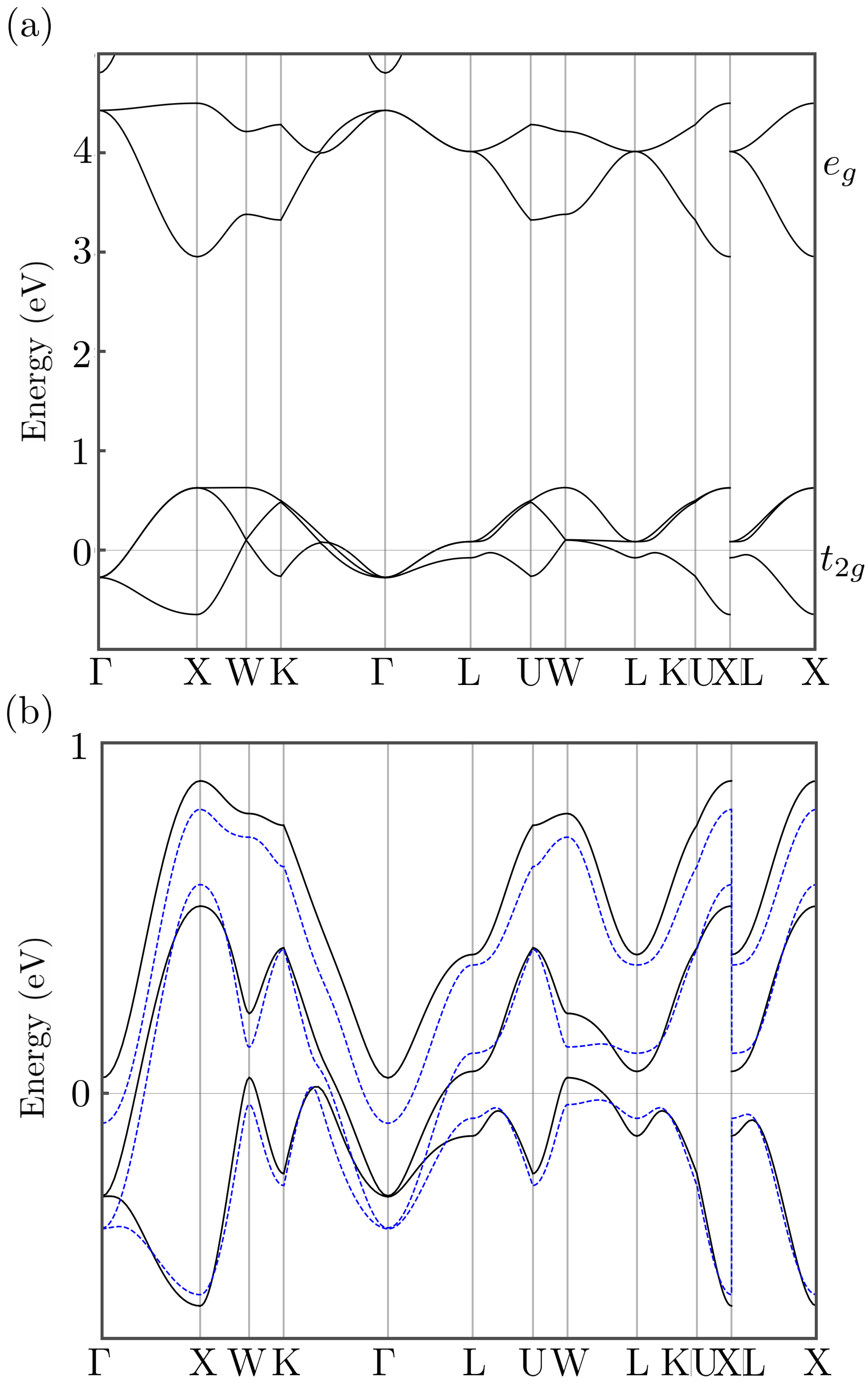}
    \caption{(a) Band structure of Ba$_2$MgOsO$_6$ computed using GGA without SOC. (b) Solid black lines represent the band structure of Ba$_2$MgOsO$_6$ around the Fermi energy computed using GGA+SOC. 
    Dashed blue lines denote the tight-binding bands with $\xi = 0.25$ eV and tight-binding parameters obtained using the MLWF derived from DFT for Ba$_2$MgOsO$_6$.
    \label{DFTband}} 
\end{figure}

In this section, we will use DFT to estimate values for the octahedra crystal field splitting ($\Delta$), atomic SOC ($\xi$), and tight-binding parameters 
($t_1, t_2, t_3$) for Ba$_2$XOsO$_6$ (X = Mg, Ca).
They determine $J_\tau$, $J_q$, and $J_o$, which will be then used to obtain the classical ground-state order for these materials. 
{ Since there has been no observed distortions in these materials, we set $t_4=0$ to represent an ideal structure.}
Here we show the results for Ba$_2$MgOsO$_6$, and similar results are obtained for {Ba$_2$BOsO$_6$ (B = Cd, Ca)}.

The band structures obtained by GGA and GGA+SOC are presented in Fig. \ref{DFTband}(a) and (b), respectively, where the Perdew-Burke-Ernzerhof functional\cite{Perdew2008PRL,Perdew2009PRL} and an 8$\times$8$\times$8 k-grid are used. The bands well below the Fermi energy are dominated by contributions from Ba, Mg and O, while the bands around the Fermi energy 
mainly arise from the Os atoms. The orbital composition of the bands near the Fermi energy are dominated by the $t_\textrm{2g}$ orbitals
and the $e_\textrm{g}$ lie around 4 eV, giving us an estimation for the octahedra crystal field splitting, $\Delta \sim 4$ eV. 
We also determine the tight binding parameters using maximally localized Wannier functions (MLWF) generated from OpenMX.\cite{Neale2016,Pritikin2015,Hunter2018}

Fig. \ref{DFTband}(b) shows how $t_{2g}$ bands near the Fermi energy are modified by the finite SOC. The black solid line represents the band structure obtained by GGA + SOC. 
By fitting the n.n. tight-binding bands to the GGA+SOC bands from DFT, we estimate $\xi \approx 0.25$ eV.  The blue dashed line in Fig. 2(b) represents
the band structure obtained by the tight binding parameters (listed in Table 1) with $\xi = 0.25$ eV.

We also estimate the splitting between the doublet and triplet $\Delta_c$.
Taking $U=2.5$ eV and $J_H=0.25$ eV, which are typical for 5d transition metal materials\cite{rau2016review,Bo2017PRB} and
$\Delta= 4$ eV,  
we find the non-Kramers doublet and triplet splitting $\Delta_c \sim$ 22 meV by numerically diagonalizing Eq. \ref{kanamoriHam} including the $e_g$ orbitals.
This estimation can be compared to recent inelastic neutron scattering results which suggests $\Delta_c$ is approximately 10-20 meV\cite{Maharaj2020PRL}.

{ Finally, $J_\tau$, $J_q$ and $J_o$, are shown in Table I.} 
Notice that $J_\tau^{(2)}$, which contributes to the bond-dependent quadrupolar interactions, 
is the dominant interaction which is approximately four times larger than the bond-independent FM octupolar and FM quadrupolar interactions. 
{ $J_\tau^{(4)}$ shows the suppression of the quadrupolar interaction by taking into account fourth order processes via the triplet states (see Table I); this brings these systems 
much closer to the FM octupolar phase boundary (see Fig. 6). The suppression of the quadrupolar interaction in Ba$_2$MgOsO$_6$ is large enough to push this compound across the phase boundary 
from the AFM type-I quadrupole phase to a FM octupole phase (Fig. 6). Due to the reduced $t_3$ in Ba$_2$BOsO$_6$ (B = Ca, Cd), 
the suppression of the quadrupolar interaction is not as large and both of 
these compounds remain in the AFM type-I quadrupole phase (Fig. 6).
} 
{\renewcommand{\arraystretch}{1.7}
\setlength{\tabcolsep}{0.1cm}
\begin{center}
    \begin{tabular}{||c || c  c  c  || c c c || c c c ||} 
        \hline
        B   & $t_3$ & $t_2$ & $t_1$  &  $J_\tau^{(2)}$ & $J_q^{(2)}$  & $J_o^{(2)}$ &  $J_\tau^{(4)}$ & $J_q^{(4)}$  & $J_o^{(4)}$\\ [0.5ex] 
        \hline\hline
        Mg  & -140 & 19.1 & 17.2 & 4.4 & -1.3 & -1.1 & 2.8 & -1.5 & -1.1\\
        \hline
        Ca & -125 & 16.9 & 13.6 & 3.4 & -0.93 & -0.78 & 2.4 & -1.0 & -0.8 \\
        \hline
        Cd & -88.8 & 17.5 & 13.6 & 1.9 & -0.68& -0.51 & 1.6 & -0.7 & -0.5 \\
        \hline
    \end{tabular}
    \captionof{table}{Hopping integral and exchange integral energies for Ba$_2$BOsO$_6$ (B = Mg, Ca, Cd). All values are written in meV. { The exchange integrals $J_\tau$, $J_q$ and $J_o$
    are denoted by a superscript (2) or (4) to denote if they were computed using second or fourth order perturbation theory respectively.}} 
\end{center}
}

\begin{figure}
    \includegraphics[width=0.4\textwidth]{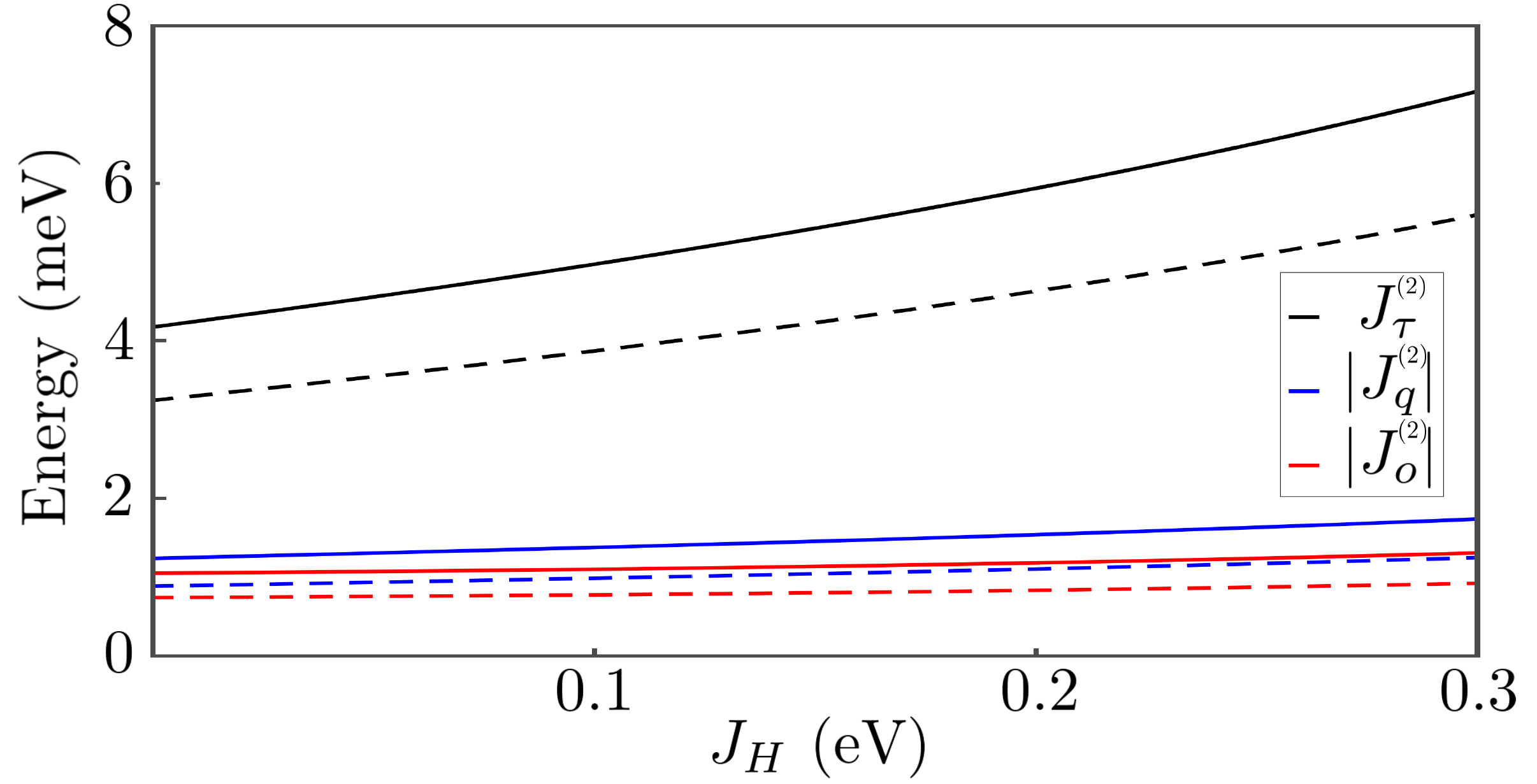}
    \caption{Exchange parameters for Ba$_2$BOsO$_6$ as a function of $J_H$ with fixed $\xi=0.25$ eV. Note that we plot $|J_o^{(2)}|$ and $|J_q^{(2)}|$.
    The solid and dashed lines denote B=Mg and Ca, respectively.
    \label{Hunds}}
\end{figure}

These exchange parameters depend on the Hund's coupling, even though we omitted it in the analytical expression in Eq. \ref{jt}.
To show its impact on them, we show their dependence on $J_H$ in Fig. \ref{Hunds}. Note that AFM $J_\tau$ becomes stronger as $J_H$ increases.
This will be discussed in the following section when we present the classical ground state using the exchange parameters in Table 1.

\section{Classical Monte-Carlo Simulations}

In this section, we determine the classical ground state order of our spin model using Monte Carlo simulations and give an estimate for the transition temperature. We use a classical Monte-Carlo algorithm known as simulated annealing; our simulated annealing code is based on the framework provided by the ALPS project \cite{Albuquerque2007,Bauer_2011,Troyer1998}. We used a $N=1728$ site cluster (12$\times$12$\times$12 primitive unit cells) with periodic boundary conditions. Once the system thermalized 
at a temperature of interest, $10^5$ measurements were acquired with 500 sweeps between each measurement. 

\subsection{Exchange integrals without triplet contributions}

\begin{figure}[H]
    \includegraphics[width=0.48\textwidth]{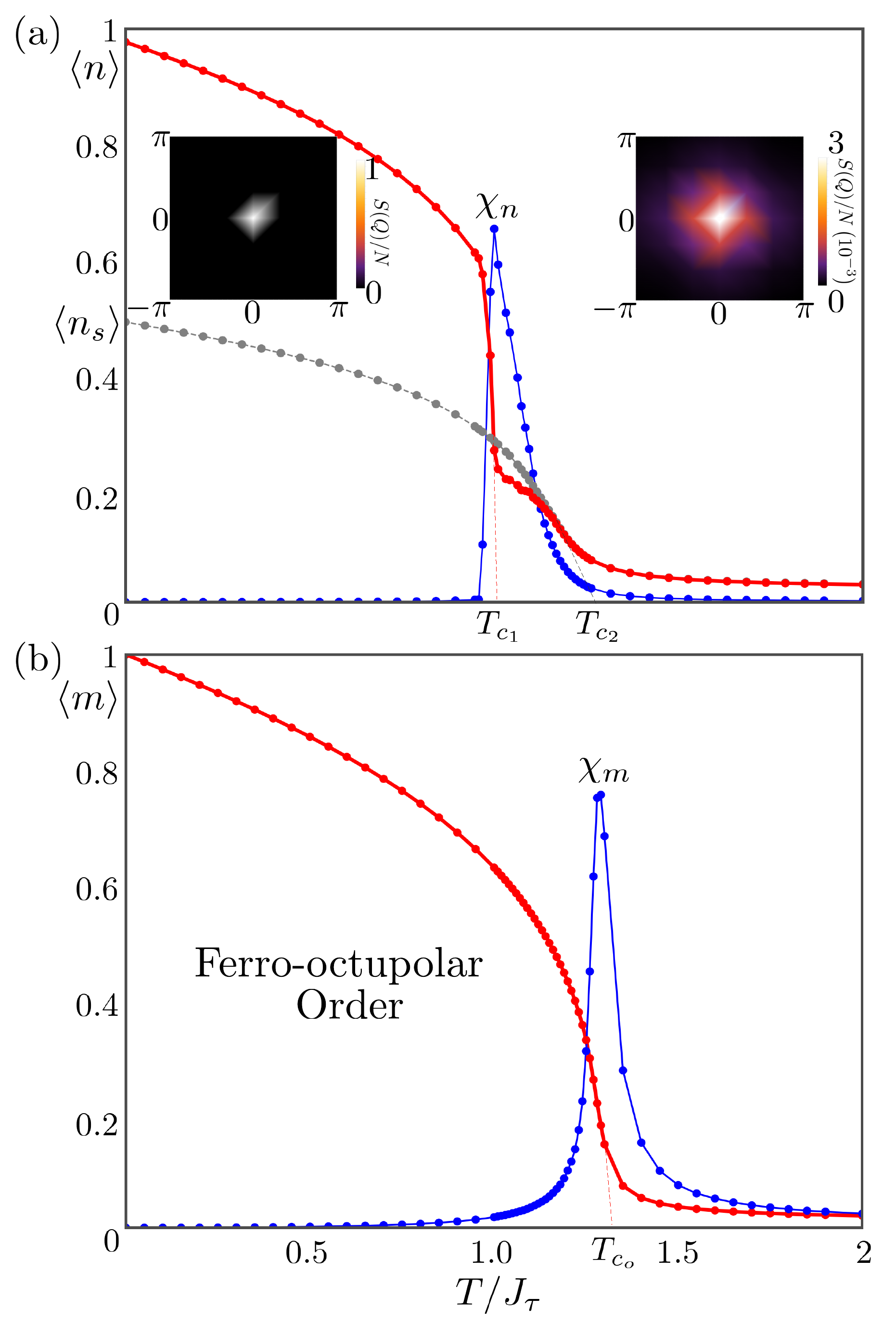}
   \caption{(a) AFM Type-I order parameter $\langle n \rangle$ and the susceptibility $\chi_n $ as a function of temperature $\left(T\right)$ with $J_\tau =1$ are shown by the red and blue lines,
    respectively. The other order parameter measured by the correlation among the same sublattice $\langle n_s \rangle$ is shown by the grey line, indicating the same sublattice correlation appears at $T_{c_2} \sim1.25 J_\tau$, while the long-range type-I AFM occurs only below $T_{c_1} \sim 1.07 J_\tau$. The static structure factors $S({\bf q})$ at fixed $q_z = 2\pi$ in the AFM ordered and partial ordered phases are shown in the left and right insets, respectively. 
   The maximum intensity of $S({\bf q})$ in the partial order phase is $3 \times 10^{-3}$. { This plot was computed using $J_\tau^{(2)}$, $J_q^{(2)}$ and $J_o^{(2)}$ for Ba$_2$MgOsO$_6$. (b) 
   FM order parameter $\expval{m}$ and the susceptibility $\chi_m$ as a function of temperature $(T)$ with $J_\tau = 1$ are shown by red and blue lines respectively. 
   This plot was computed including triplet processes (using $J_\tau^{(4)}$, $J_q^{(4)}$ and $J_o^{(4)}$) for Ba$_2$MgOsO$_6$.}
   \label{OP}} 
\end{figure}

{The resulting order for Ba$_2$BOsO$_6$ (B = Mg, Ca, Cd), using the exchange parameters obtained from second order perturbation theory ($J_\tau^{(2)}$, $J_q^{(2)}$, and $J_o^{(2)}$), is quadrupolar AFM type-I order.}
The order parameter is measured by the thermal average $\langle n \rangle$ where $n =\sqrt{\sum_{ij} e^{i{\bf q} \cdot ({\bf r_i} - {\bf r_ j}) } {\bf s}_i \cdot {\bf s}_j}$ 
where
${\bf q}= (0,0, 2 \pi/a)$. 
A plot of order parameter denoted by the red line as a function of temperature is shown in Fig. \ref{OP}(a). There is a sharp jump at the transition temperature $T_{c1} = 1.07 J_\tau$ indicating a first order transition. 
The susceptibility is measured by $\chi_n \propto \expval{n^2} - \expval{n}^2 $ denoted by the blue line which shows a peak at the transition temperature.
This order is expected since the quadrupolar and octupolar FM terms are approximately 4 times smaller than the $J_\tau$ term which we have shown in an earlier work, carries a quadrupolar AFM type-I order on the FCC lattice \cite{Kha2021PRR}. This order is also observed using the exchange interactions at a finite Hund's coupling ($J_H=0.25$ eV) and fixed SOC ($\xi=0.25$ eV).

Surprisingly, there is an additional shoulder above $T_{c_1}$ which eventually disappears above $T_{c_2} (\sim 1.25 J_\tau)$.
To understand the nature of the shoulder, we compute the quadrupole-quadruple correlation among moments within the same sublattice, 
i.e., $\langle n_s \rangle = \langle \sqrt{\sum_{ij \in A}{\bf s}_{i} \cdot {\bf s}_j} \rangle$ where ${\bf s} = s_x  {\hat x} + s_z {\hat z}$ and $i$ and $j$ belong to a same sublattice $A$.
Its associated order parameter is shown in the grey line $\langle n_s  \rangle$
as shown in Fig. \ref{OP}(a). This implies that there is a partial order, where the stripy quadrupole ordered pattern is lost
within the unit cell, while keeping the long-range order between unit cells. 
This can be contrasted with the pure $J_\tau$ model (with $J_o,J_q=0$), which only undergoes a single first order phase transition.\cite{Kha2021PRR}

The static structure factors $S({\bf q})$ for fixed $q_z = 2 \pi$  in the type-I AF at $T=0$ and partial ordered phases at $T= 1.07 J_\tau$ are also plotted in the inset of Fig. \ref{OP}(a); 
we set $a \equiv 1$, the side length of FCC unit cell.
  As expected, there is a sharp delta-function feature at $(0,0, 2\pi)$ inside the type-I AFM order and the moments are all in the plane perpendicular to 
  the [111]-axis implying the quadrupolar order. On the other hand, the static structure factor inside the partial order shows the blur feature
  maximized around $(0,0, 2\pi)$ with the maximum intensity of $3 \times 10^{-3}$  indicating type-I  AFM order is lost.

\begin{figure}
    \includegraphics[width=0.4\textwidth]{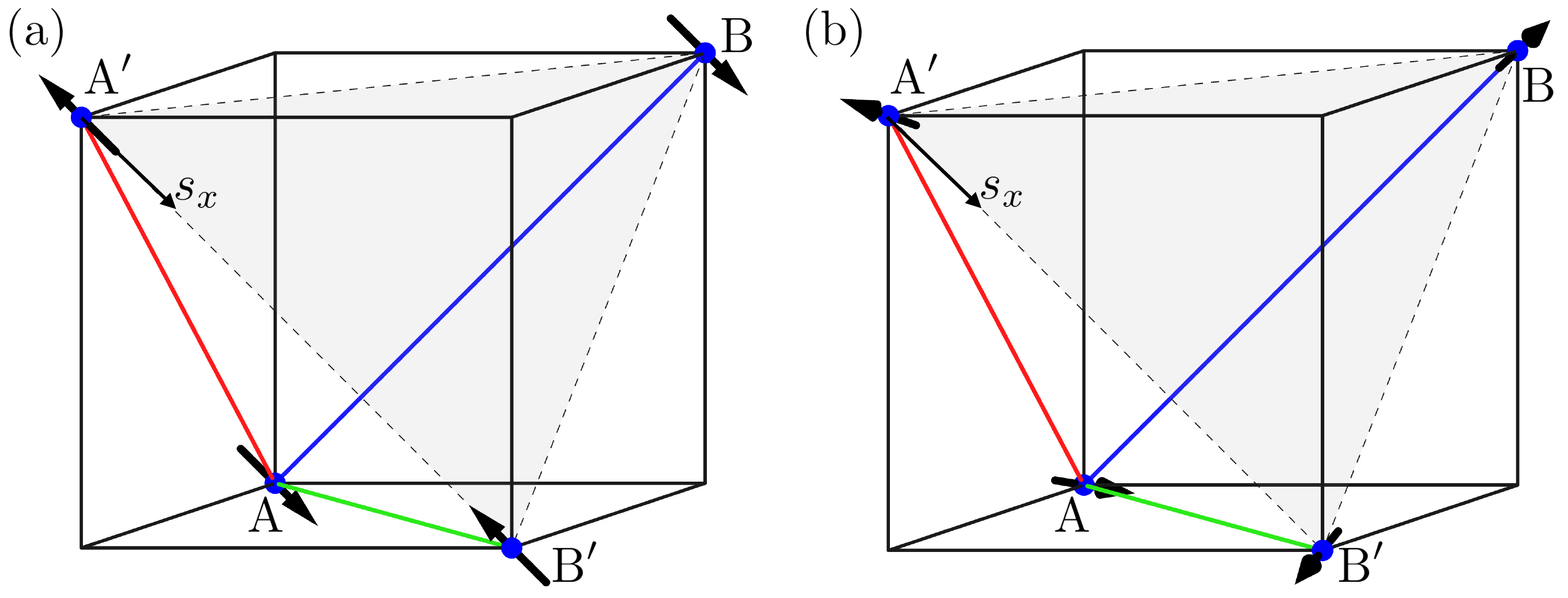}
    \caption{(a) Stripy quadrupole order and (b) partially ordered quadrupolar state at $T = 1.07 J_\tau$, with exchange integral energies equal to those in Table 1 for Ba$_2$MgOsO$_6$. 
    A, A$'$, B, B$'$ label the four different moments within FM sublattices, and the psuedospin x-axis is denoted $s_x$. 
    The same phase is observed for Ba$_2$CaOsO$_6$, in similar temperature regime.
    \label{DPorder}}
\end{figure}

\begin{figure}
    \includegraphics[width=0.4\textwidth]{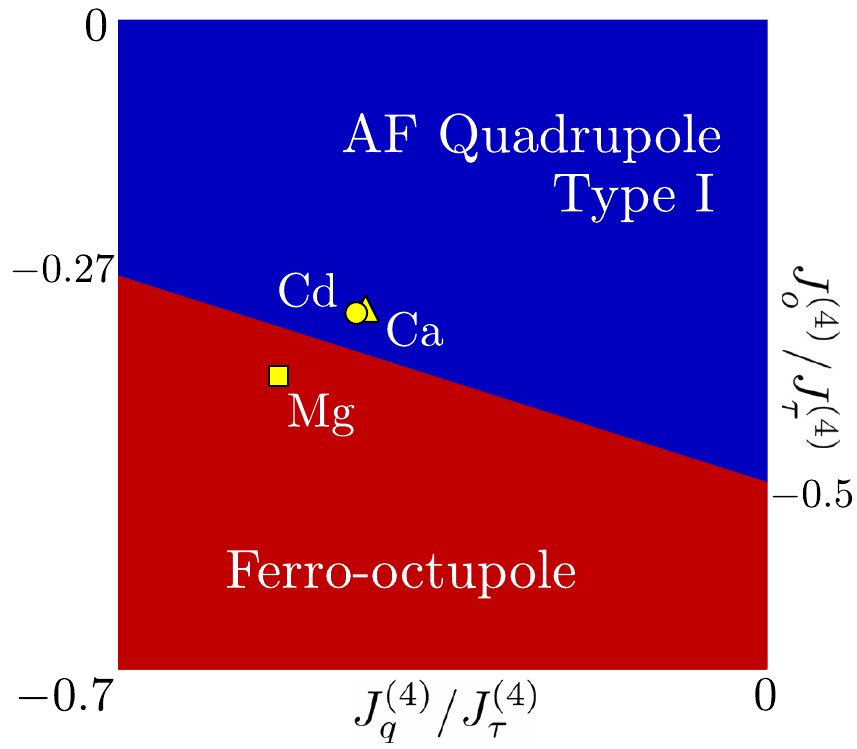}
    \caption{{Zero temperature phase diagram of Eq. \ref{tot_ham} in a parameter regime relevant for Ba$_2$BOsO$_6$ (B = Mg, Ca, Cd). The square, triangle and circle correspond to 
    the exchange integrals including fourth order triplet processes ($J_\tau^{(4)}$, $J_q^{(4)}$, and $J_o^{(4)}$) for Ba$_2$MgOsO$_6$, Ba$_2$CaOsO$_6$ and Ba$_2$CdOsO$_6$ respectively.}
    \label{DPpd}}
\end{figure}

To understand the nature of the partial order, we compute 
the averaged quadrupole moment in this phase, which appears
immediately after first phase transition  $T_{c_1}$ and below $T_{c_2}$.  The averaged moment is shown in Fig. \ref{DPorder}. Due to
thermal fluctuations, the moment fluctuates widely within the plane, but it has a finite moment on average. 
The new partially ordered phase has four different moments inside the FCC unit cell, which then repeat forming FM ordering among 
the same sublattices.
This order along with the sublattices (A, A$'$, B, B$'$) are shown in Fig. \ref{DPorder}. 
This result was obtained using classical Monte Carlo with a 1372 site cluster (7x7x7 conventional unit cells each with 4 sites) at $T = 1.07 J_\tau$.
This figure shows the average spin configuration over 10$^5$ sweeps.   
All pseudospins lie within the [111] plane, and thus this order is purely quadrupolar. 
On average, moments on the A sublattice make a 42$^\circ$ angle with the $s_x$ axis, 
and the angle between moments on the A and A$'$ is 190$^\circ$. 
Moments on the B sublattice make a 98$^\circ$ angle with the $s_x$ axis, and the angle between moments on the B and B$'$ is 173$^\circ$.

\subsection{Exchange integrals including triplet contributions}
{
The zero temperature phase diagram computed with CMC, using the exchange integrals including the triplet processes which appears at fourth order ($J_\tau^{(4)}$, $J_q^{(4)}$, and $J_o^{(4)}$), is shown in Fig. \ref{DPpd}. 
The resulting order for Ba$_2$MgOsO$_6$ is FM octupolar order; consistent with other recent 
studies on these materials \cite{Pouro2021PRL, Voleti2021PRB}. The FM order parameter is given by the thermal average $\expval{m}$, where
$m =\sqrt{\sum_{ij} {\bf s}_i \cdot {\bf s}_j}$, and is plotted in red along with its susceptibility $\chi_m \propto \expval{m^2} - \expval{m}^2$ in blue (Fig. \ref{OP}(b)).
The long-range FM octupolar order sets in at temperatures below $T_{c_o} \sim 1.4 J_\tau$. 
Note this order is stabilized at a higher temperature than the AFM Type-I quadrupolar order because FM interactions are not frusturated on 
the FCC lattice and are thus more resistant to thermal fluctuations.

However both Ba$_2$BOsO$_6$ (B = Ca, Cd), even including triplet contributions which suppresses the quadrupolar interactions, exhibit AFM Type-I quadrupole groundstates at zero temperature; this
can be contrasted to \cite{Pouro2021PRL, Voleti2021PRB} which predict a more negative $J_o$, and FO octupolar order for Ba$_2$CaOsO$_6$. This discrepancy is likely caused by
our result only including processes via the triplet up to fourth order. Moreover, \cite{Pouro2021PRL} uses a smaller value for $\Delta_c$ for Ba$_2$CaOsO$_6$ which will further suppress our 
quadrupolar interactions and bring Ba$_2$CaOsO$_6$ even closer to the FM octupolar boundary.
}

\section{Summary and Discussion}
In summary, we find 
that the \textit{interference} of two intra-orbital hopping processes generates FM octupolar interactions, which dominates over the other contributions leading to
AFM interaction, and the overall octupolar interaction is thus FM type.
This exchange process also contributes to the bond-independent quadrupolar interactions, which competes with the FM octupole order.
The origin of such bond-dependent and -independent exchange interactions can be traced back to the shape of the doublet wavefunctions. 
Using {\it ab-initio} calculations, we determine the SOC and tight binding parameters which in turn determine 
the strengths of the exchange interactions of the pseudospin model.
For Ba$_2$BOsO$_6$ where B = Mg, Cd, Ca, we find FM octupolar interactions together with FM bond-independent and AFM bond-dependent
quadrupolar interactions. 

We used classical Monte Carlo simulations to determine the classical ground state for these double perovskites. When processes via the triplet were ignored, we
 found the AFM type-I quadrupolar order with an ordering temperature to be approximately $T_{c_1} \sim 50 K$, despite octupolar FM exchange interactions.  Just above $T_{c_1}$, there is a partial quadrupolar order between $T_{c_1}$ and $T_{c_2}$
 where the stripy pattern within the unite cell is lost, while a long-range correlation among the same sublattice, i.e, FM sublattice 
is preserved. {
When exchange processes via the triplet are considered using fourth order perturbation theory,
we find FM octupolar order in Ba$_2$MgOsO$_6$ and that AFM type-I quadrupolar order persists in Ba$_2$BOsO$_6$ (B = Ca, Cd).
}

We find that octupolar order can be achieved with $J_q =-0.30 J_\tau$ (like in the case for Ba$_2$MgOsO$_6$), when $J_o < -0.41J_\tau$. 
If one can reduce $J_q$ slightly, the threshold to achieve octupolar order is moved closer to the value of $J_o$ for 
Ba$_2$MgOsO$_6$. 
{ This implies that these materials exist in a parameter regime close to octupolar FM order. However, this also suggests that coupling to the lattice through distortions 
may become significant and amplify the quadrupolar interactions via Jahn-Teller coupling.}
Its relation to
the internal magnetic field reported by $\mu SR$ measurements is a puzzle for future study.
Due to strong SOC, the coupling to the lattice would be strong, and quantifying such effects remain to be studied further. 
Another interesting direction is designing new materials exhibiting an intriguing pattern of vortex quadrupole and ferri-octuploar order\cite{Kha2021PRR}.
Theoretically this can be achieved by tuning inter-orbital $t_1$ to be negligible, while enhancing inter-orbital $t_2$. Synthesizing such new material
is an excellent project for future studies.
 
\begin{acknowledgments}
We would like to thank G. Khaliullin, P. Stavropoulos, B. Gaulin for useful discussions.
H.Y.K. acknowledges support from the NSERC Discovery Grant No. 06089-2016, and support from CIFAR
and the Canada Research Chairs Program. 
Computations were performed on the Niagara supercomputer at
the SciNet HPC Consortium. SciNet is funded by: the
Canada Foundation for Innovation under the auspices of
Compute Canada; the Government of Ontario; Ontario
Research Fund - Research Excellence; and the University
of Toronto.
\end{acknowledgments}

\bibliography{d2_dp_multipoles_aug26}

\appendix 
{
\section{Fourth order perturbation theory including virtual processes via triplet states}
At fourth order perturbation theory, the dominant processes are those which connect the non-Kramers doublet to the excited $T_{2g}$ triplet. 
Processes via the triplet simultaneously suppresses the AFM $J_\tau$ term and enhances the FM $J_o$ term. The exchange integrals including triplet processes are shown below. 
Notice these reduce to $J_\tau^{(2)}$, $J_q^{(2)}$, $J_o^{(2)}$ when the fourth order processes are removed. 
\begin{align}\label{exintcor}
    {J_\tau^{(4)}}&=\frac{4 {t_3}^2}{9 U}\biggl[1 -\frac{3 {t_3} ({t_3}-{t_1})+ 2 {t_2}^2}{3 U \Delta_c} \\
    &\notag -\frac{2 \left({t_3}\left(7 {t_3} -22 {t_1}\right) +15 {t_2}^2 + 33 {t_1}^2\right)}{9 U^2}\biggr] \\
    &\notag -\frac{8 {t_1} {t_3}}{9 U}\biggl(1-\frac{4 {t_2}^2 + 3 {t_1}^2}{6 U \Delta_c}-\frac{30 {t_2}^2 + 28 {t_1}^2}{9 U^2}\biggr) \\
    &\notag +\frac{4 {t_1}^2}{9 U}\biggl(1-\frac{11 {t_2}^2 + 3 {t_1}^2}{3 U\Delta_c}-\frac{30 {t_2}^2 + 20 {t_1}^2}{9 U^2}\biggr) \\
    &\notag -\frac{2 {t_2}^4}{3 U^2 \Delta_c } \\
    \notag {J_q^{(4)}} &= \frac{2t_1}{3U}\biggl[t_1\biggl(1- \frac{12 t{_3} \left(3 {t_3} +2 {t_1}\right) + 33 {t_1}^2 +10 {t_2}^2}{24 U \Delta_c } \\
    &\notag -\frac{18 t_3 \left({t_3} +4{t_1}\right) + 36 {t_1}^2+4 {t_2}^2}{9 U^2}\biggr) \\
    &\notag +{2{t_3}}\biggl(1-\frac{3 {t_3}^2 + 2 {t_2}^2}{12 U \Delta_c}-\frac{18 {t_3}^2 + 32 {t_2}^2}{9 U^2}\biggr)\biggr] \\
    &\notag -\frac{2 {t_2}^2}{3 U}\biggl(1+ \frac{2 {t_3}^2 + 7 {t_2}^2}{8 U \Delta_c }-\frac{14 {t_3}^2 + 36 {t_2}^2}{9 U^2}\biggr) \\
    &\notag -\frac{{t_3}^4}{12 U^2 \Delta_c} \\
    \notag {J_o^{(4)}} &= \frac{2 {t_1}}{3 U} \biggl[t_1\biggl(1+\frac{3 t_3\left( {t_3} -2 {t_1}\right)+68 {t_2}^2 + 3 {t_1}^2}{12 U \Delta_c} \\
    &\notag -\frac{18 t_3 \left(t_3 + 4t_1\right) + 36 t_1^2 + 56 t_2^2}{9 U^2}\biggr) \\
    &\notag +2t_3 \biggl(1+\frac{11 {t_2}^2}{12 U \Delta_c}-\frac{18 {t_3}^2 + 28 {t_2}^2}{9 U^2}\biggr)\biggr] \\
    &\notag +\frac{2 {t_2}^2}{3 U}\biggl(1 + \frac{2 {t_3}^2 + 3 {t_2}^2}{4 U \Delta_c }-\frac{14 {t_3}^2 + 36 {t_2}^2}{9 U^2}\biggr)
\end{align}
}

\end{document}